# The superconducting ferromagnet UCoGe

A. Gasparini[1], Y.K. Huang[1], N.T. Huy[1,*], J.C.P. Klaasse[1], T. Naka[2], E. Slooten[1] and A. de Visser[1]

[1]*Van der Waals-Zeeman Institute, University of Amsterdam, Valckenierstraat 65, 1018 XE Amsterdam, The Netherlands*
[2]*National Research Institute for Materials Science, Sengen 1-2-1, Tsukuba, Ibaraki 305-0047, Japan*

*The correlated metal UCoGe is a weak itinerant ferromagnet with a Curie temperature $T_C = 3$ K and a superconductor with a transition temperature $T_s = 0.6$ K. We review its basic thermal, magnetic − on the macro and microscopic scale − and transport properties, as well as the response to high pressure. The data unambiguously show that superconductivity and ferromagnetism coexist below $T_s = 0.6$ K and are carried by the same 5f electrons. We present evidence that UCoGe is a p-wave superconductor and argue that superconductivity is mediated by critical ferromagnetic spin fluctuations.*

Corresponding author:
Dr. A. de Visser
Van der Waals-Zeeman Institute, University of Amsterdam
Valckenierstraat 65, 1018 XE Amsterdam, The Netherlands
E-mail: a.devisser@uva.nl

---

* Current address: Hanoi Advanced School of Science and Technology, Hanoi, University of Technology, 1 Dai Co Viet, Hanoi, Vietnam.

## 1. Introduction

The intermetallic compound UCoGe belongs to the intriguing family of superconducting ferromagnets [1,2]. In superconducting ferromagnets, a superconducting transition takes place at a temperature $T_s$ deep in the ferromagnetic state, *i.e.* well below the Curie temperature $T_C$, without expelling magnetic order. The superconducting ferromagnets discovered hitherto are UGe$_2$ (under pressure) [3], URhGe [4], UIr (under pressure) [5], and UCoGe. In these uranium intermetallics magnetism has a strong itinerant character and both ordering phenomena are carried by the same $5f$ electrons. The coexistence of superconductivity and ferromagnetism is at odds with the standard BCS theory for phonon-mediated $s$-wave superconductivity, because the ferromagnetic exchange field is expected to inhibit spin-singlet Cooper pairing [6]. The band nature of the ferromagnetic order, however, allows for an alternative explanation, in which critical spin fluctuations provide the mechanism for pairing up the spin-split band electrons in spin-triplet Cooper states [7,8]. In recent years ample evidence has been presented that such an unusual pairing mechanism is at work in superconducting ferromagnets [3,9,10,11].

With the discovery of superconducting ferromagnets a new research theme in the field of magnetism and superconductivity has been disclosed. Research into ferromagnetic superconductors will help to unravel how magnetic fluctuations can stimulate superconductivity, which is a central theme running through materials families as diverse as the heavy-fermion superconductors, high-$T_s$ cuprates and the recently-discovered FeAs-based superconductors [12]. This novel insight might turn out to be crucial in the design of new superconducting materials.

The coexistence of superconductivity and weak itinerant ferromagnetism in UCoGe was reported in 2007 [1]. Till then UCoGe was thought to be a paramagnet down to a temperature of 1.2 K [13]. However, in a search for a ferromagnetic quantum critical point induced in the superconducting ferromagnet URhGe ($T_s$ = 0.25 K, $T_C$ = 9.5 K) by alloying with Co [14], it was discovered that UCoGe is actually a weak itinerant ferromagnet below $T_C$ = 3 K and, moreover, a superconductor below $T_s$ = 0.8 K.

In this paper we review the basic thermal, magnetic and transport properties of UCoGe. Magnetization measurements show that UCoGe is a uniaxial ferromagnet, and that the ordered moment $m_0$ = 0.07 $\mu_B$ is directed along the orthorhombic $c$ axis [15]. Muon spin relaxation measurements [16] provide unambiguous proof that magnetism is a bulk property, which coexists with superconductivity on the microscopic scale. The temperature variation of the upper critical field $B_{c2}(T)$ [15] shows an unusual upward curvature and is not Pauli limited for $B \parallel a$ and $B \parallel b$, which provides solid evidence for spin-triplet Cooper pairing. High-pressure susceptibility and transport experiments [17] reveal that ferromagnetic order is smoothly depressed and vanishes at a critical pressure $p_c \approx 1.4$ GPa. Near the ferromagnetic critical point superconductivity is enhanced, which yields strong support for superconductivity stimulated by critical ferromagnetic spin fluctuations.

## 2. Metallurgical aspects and sample preparation

UCoGe crystallizes in the orthorhombic TiNiSi structure (space group $P_{nma}$) [18], with room-temperature lattice parameters $a = 6.845$ Å, $b = 4.206$ Å and $c = 7.222$ Å [1]. Superconductivity and magnetic order were first observed on annealed polycrystalline samples with nominal compositions $U_{1.02}CoGe$ (sample #2, $RRR \approx 10$) and $U_{1.02}Co_{1.02}U$ (sample #3, $RRR \approx 30$) prepared by arc melting [1]. Here $RRR = R(300K)/R(1K)$ is the residual resistance ratio. The coexistence of superconductivity and ferromagnetism is a robust property of all polycrystalline samples subjected to an appropriate heat treatment procedure, typically a period of ten days at a temperature of 875 °C.

Single-crystalline samples were pulled from the melt with nominal composition $U_{1.01}CoGe$ using a modified Czochralski technique in a tri-arc furnace under a high-purity argon atmosphere [19]. To improve the sample quality, pieces of the single crystals, cut by spark-erosion, were annealed in evacuated quartz tubes for one day at 1250 ºC and 21 days at 880 ºC. This heat-treatment procedure is similar to the one applied to URhGe [20] and led to a significant increase of the $RRR$ value from 5 to ~30. The still relatively low $RRR$ value is possibly caused by remaining disorder due to Co and Ge site inversion. Notice, the TiNiSi structure is an ordered variant of the $CeCu_2$ structure, in which Co and Ge atoms randomly occupy the 4$c$ positions [21].

The temperature dependence of the resistivity of annealed single-crystalline UCoGe for a current along the $b$ axis is shown in Fig. 1. The $RRR$ value amounts to 40. Proper superconducting and ferromagnetic phase transitions are observed. The magnetic transition is represented by a sharp kink at $T_C = 2.8$ K, and superconductivity appears at temperatures below $T_s = 0.6$ K. However, the superconducting transition is still relatively broad, $\Delta T_s \approx 0.06$ K.

## 3. Weak itinerant ferromagnetic order

Magnetization data taken on polycrystalline samples provide solid evidence that UCoGe is a weak itinerant ferromagnet, with a Curie temperature $T_C$

= 3 K as deduced from Arrott plots [1]. A hysteresis loop with a coercive field of ~0.3 mT at $T$ = 2 K corroborates ferromagnetic order. The polycrystalline averaged ordered moment amounts to $m_0$ = 0.03 $\mu_B$ for $T \to 0$. Consequently, the ratio, $p_{eff}/M_S$, of the Curie-Weiss effective moment $p_{eff}$ = 1.7 $\mu_B$ over the saturation moment $M_S$ is small, which classifies UCoGe as a weak itinerant ferromagnet [22].

Magnetization measurements on a single-crystalline sample show UCoGe is a uniaxial ferromagnet [15]. The field dependence of the magnetization $M(B)$ measured in fields up to 5 T applied along the $a$, $b$ and $c$ axis at a temperature of 2 K is shown in Fig. 2. In the inset we show the temperature variation $M(T)$ measured in a field $B \parallel c$ of 0.01 T. The Curie temperature $T_C$ = 2.8 K is determined by the inflection point in $M(T)$. $M(T)$ is well expressed by the relation $M(T) = m_0(1 - (T/T^*)^2)^{1/2}$ [23], predicted for weak itinerant ferromagnets [24], with $T^* \sim T_C$ and the ordered moment $m_0 \approx 0.07$ $\mu_B$/f.u.

Resistivity measurements on a single-crystalline sample show the magnetic transition is presented by a sharp kink at $T_C$ = 2.8 K (see Fig. 1). In the temperature ranges below and above $T_C$ the resistivity follows the typical Fermi-liquid $\rho \sim T^2$ and $\sim T^{5/3}$ laws [24], respectively. The temperature exponent $n = 5/3$ is characteristic for scattering at critical ferromagnetic spin fluctuations. Transport measurements in a magnetic field $B \parallel c$ reveal the ferromagnetic transition is rapidly smeared out (see Fig. 1).

The thermodynamic signature of the ferromagnetic transition in the specific heat measured on a polycrystalline sample (labelled sample #2) is shown in Fig. 3. Here $T_C$ = 3 K is identified by the inflection point in $c/T$ at the high $T$ side of the peak. The linear term in the electronic specific heat $\gamma$ amounts to 0.057 J/molK$^2$, which indicates UCoGe is a correlated metal, but the electron interactions are relatively weak. The magnetic entropy $S_{mag}$ involved in the magnetic transition, obtained by integrating $c_{mag}/T$ versus $T$, is 0.3 % of $R\ln2$ (*i.e.* the value for a local moment $S = ½$ system). Such a small value is expected for a weak itinerant ferromagnet [24]. In small applied magnetic fields ($B \leq 0.3$ T) the magnetic entropy is preserved, but the ferromagnetic transition broadens significantly, as shown in Fig. 3.

### 4. Superconductivity

Typical ac-susceptibility, $\chi_{ac}$, traces taken on polycrystalline UCoGe samples are shown in Fig. 4. These data were taken at a low frequency ($f$ = 16 Hz) and an amplitude of the driving field of ~10$^{-5}$ T. The weak peak observed at 3 K signals the ferromagnetic transition. Below 1 K, $\chi_{ac}$ rapidly decreases to a large diamagnetic value, which reflects the superconducting transition. The onset transition temperatures, $T_s^{onset}$, are determined at 0.38 K and 0.61 K for samples #2 ($RRR \approx 10$) and #3 ($RRR \approx 30$), respectively. Clearly, superconductivity depends sensitively on the quality of the samples. These results are in good agreement with resistivity data taken on the same samples [1,23]. The ac-susceptibility $\chi_{ac}$ starts to drop when the resistive transition is complete. At the lowest temperature $\chi_{ac}$ reaches a value of 60–70 % of the ideal screening value $\chi_S = -1/(1 - N)$ (here $N \approx 0.08$ is the demagnetizing factor of our samples). This indicates UCoGe is a Type II superconductor which is always in the mixed phase. Because of the intrinsic ferromagnetic moment, the local field is nonzero and the magnitude of $\chi_{ac}$ is reduced.

The temperature dependence of the ac-susceptibility of sample #3 in magnetic field is shown in the inset of Fig. 4. In an applied field, the peak associated with the ferromagnetic order broadens and shifts to higher temperatures, while the onset temperature for superconductivity shifts to lower temperatures. In a field of 0.1 T, the magnetic transition in $\chi_{ac}(T)$ is almost washed out.

Specific-heat and thermal-expansion measurements provide solid evidence for bulk superconductivity. Specific-heat data taken on a polycrystalline sample show a broad superconducting transition with an onset temperature of 0.66 K [1]. A rough estimate for the step size of the idealized transition in the specific heat, using an equal entropy construction (with a bulk $T_s \approx 0.45$ K), yields $\Delta(c/T_s)/\gamma \approx 1.0$, which is considerably smaller than the BCS value 1.43.

The linear coefficient of thermal expansion, $\alpha = L^{-1}dL/dT$, measured on a polycrystalline sample (#3) [1] is shown in Fig. 5. Upon entering the superconducting state, $\alpha(T)$ shows a steady increase. Assuming an ideal sharp transition at a superconducting temperature $T_s = 0.45$ K, the estimated step-size $\Delta\alpha$ is $3.8\times10^{-7}$ K$^{-1}$ [1], which reflects bulk superconductivity. Moreover, the thermal expansion data reveal that magnetism and superconductivity coexist. The relative length change in the superconducting state $\Delta L/L = -0.1\times10^{-6}$ is small compared to the length change $\Delta L/L = 1.9\times10^{-6}$ due to magnetic ordering. Thus magnetism is not expelled below $T_s$ and coexists with superconductivity. Thermal expansion measurements on a single-crystalline sample for a dilatation direction along the $b$ axis show pronounced phase transition anomalies at $T_C = 2.8$ K and $T_s = 0.5$ K (see Fig. 5).

**5. Muon spin relaxation measurements**

In order to investigate whether the weak magnetic order is a bulk property of our samples, muon spin relaxation (μSR) experiments were carried out at the Paul Scherrer Institute in Villigen [16]. The experiments were carried out in zero applied magnetic field on well characterized polycrystalline samples with $RRR \approx 30$ in the temperature range 0.02 – 10 K. In the paramagnetic phase the μSR spectra are best described by a Kubo-Toyabe function $G_{KT}(t)$, with a Kubo-Toyabe relaxation rate $\Delta_{KT} = 0.30\pm0.01$ μs$^{-1}$. In this temperature range the depolarization of the muon ensemble is attributed to static nuclear moments on the $^{59}$Co atoms. In the ferromagnetic phase a clear spontaneous muon precession frequency, $\nu$, appears and the response of the muon is described by the depolarization function for an isotropic polycrystalline magnet $G_M(t)$ (see Ref.16). At the lowest temperature ($T = 0.02$ K) $\nu = 1.972\pm0.004$ MHz, which corresponds to a local field $B_{loc} \sim 0.015$ T at the muon localization site. The temperature variation $\nu(T)$ tracks the magnetization $M(T)$ (see Fig. 6). The $\nu(T)$ data are well fitted by a phenomenological order parameter function $\nu(T) = \nu_0 [1 - (T/T^*)^\alpha]^\beta$, with $T^* = 3.02$ K $\approx T_C$, $\nu_0 = 1.98$ MHz and critical exponents $\alpha = 2.3$ and $\beta = 0.4$. The amplitude of the magnetic signal in zero field below $T \sim 1.5$ K corresponds to the amplitude measured in a small transverse field of 50 G in the paramagnetic phase. This confirms magnetic order is present in the whole sample volume.

Most interestingly, $\nu(T)$ shows a small decrease of about 2% below $T_s$. This effect is observed by the whole muon ensemble, which confirms magnetism and superconductivity coexist on the microscopic scale. The decrease of $\nu(T)$ is accompanied by a small increase of the corresponding damping rate $\lambda_2(T)$ [16]. Such an increase is expected when a spontaneous vortex lattice is formed [25], *i.e.* when $B_{loc}$ is larger than the lower critical field $B_{c1}$. This indicates that even in zero magnetic field UCoGe does not enter the Meissner state, but is always in the mixed state.

### 6. Upper critical field

The upper critical field, $B_{c2}$, as measured on a single-crystalline UCoGe sample with $RRR \approx 30$ [15] is reported in Fig.7. The data were extracted from resistance, $R(T)$, measurements taken for a current along the orthorhombic $a$ axis in fixed magnetic fields applied along the $a$ axis (longitudinal configuration), and $b$ and $c$ axis (transverse configuration). The superconducting transition temperatures, $T_s(B)$, were determined by the midpoints of the transitions to the zero resistance state. In zero field $T_s = 0.6$ K.

At least three remarkable features appear in the data: (*i*) the large value of $B_{c2}^{a,b}(T=0) \approx 5$ T for $B \parallel a, b$, (*ii*) the large anisotropy, $B_{c2}^a \cong B_{c2}^b \gg B_{c2}^c$, of a factor $\sim 10$, and (*iii*) a pronounced upturn in $B_{c2}(T)$ when lowering the temperature for all field directions. Clearly, this behaviour is at odds with standard BCS spin-singlet pairing. $B_{c2}^{a,b}(T=0)$ largely exceeds the paramagnetic limiting field, $B_{c2,Pauli}(0) = 1.83T_s$, for a weak coupling spin-singlet superconductor [26]. Taking into account the 3D nature of the normal-state electronic properties of UCoGe, this provides solid evidence for an equal-spin pairing triplet superconducting state. A prerequisite for triplet pairing is a sufficiently clean sample, such that the mean free path $\ell$ is larger than the coherence length $\xi$. An estimate for $\ell$ and $\xi$ can be extracted from the large initial slope $B_{c2}^{a,b}/dT \cong -8$ T/K [27]. For our single crystal it was calculated $\ell \approx 900$ Å and $\xi \approx 120$ Å and consequently the clean-limit condition is satisfied. The large anisotropy in $B_{c2}$ has been analyzed in terms of an anisotropic $p$-wave interaction [28], which supports a superconducting gap function of axial symmetry with point nodes along the $c$ axis, i.e. along the direction of the ordered moment $m_0$. As shown in Fig. 7, $B_{c2}(T)$ is at variance with model calculations for polar and axial states of spin-triplet superconductors with a single superconducting gap function (see Ref. 15).

Superconducting order parameter calculations for orthorhombic itinerant ferromagnetic superconductors with strong spin-orbit coupling show that UCoGe is essentially a two-band superconductor [29]. Within this scenario, the unusual upward curvature at low temperatures can be attributed to a crossover between two equal-spin pairing states with different superconducting transition temperatures. Another appealing explanation for the unusual $B_{c2}$ behaviour is the presence of a field-induced quantum critical point ([30], see section 8).

### 7. Pressure – temperature phase diagram

The response to pressure of the ferromagnetic and superconducting phases of UCoGe was investigated by ac-susceptibility, $\chi_{ac}(T)$, and resistivity, $\rho(T)$, measurements on single-crystalline samples using a clamp-cell technique for pressures up to 2.2 GPa [17]. The resulting pressure-temperature phase diagram is shown in Fig. 8. The Curie temperatures obtained by both methods nicely agree. However, the values of $T_s$ determined from the $\rho(T)$ data systematically exceed those determined by $\chi_{ac}$, since the diamagnetic signal is representative for the bulk and appears when the resistive transition is complete. The Curie temperature, $T_C$, gradually decreases with pressure, and for $p \geq 0.4$ GPa a linear depression is observed at a rate $-2.4$ K/GPa. The phase line $T_C(p)$ extrapolates to the suppression of ferromagnetic order at $p_c = 1.40 \pm 0.05$ GPa. An almost equal critical pressure value is deduced from the pressure variation of the amplitude of $\chi_{ac}$ at $T_C$ (see inset Fig.8).

The susceptibility data reveal the magnetic transition is continuous – hysteresis in the magnetic signal is absent. This strongly suggests ferromagnetic order vanishes at a second order quantum critical point at $p_c$.

However, the phase line $T_C(p)$ intersect the superconducting phase boundary $T_s(p)$ near $p \approx 1.1$ GPa, above which $T_C$ no longer can be detected. Consequently, we cannot exclude that ferromagnetic order vanishes abruptly and that the ferro- to paramagnetic phase transition becomes first order when $T \to 0$ [31].

The superconducting transition temperature first increases with pressure. $T_s$ goes through a broad maximum near the critical pressure for ferromagnetic order and persists in the paramagnetic phase. This is clearly at variance with the $p$-$T$ phase diagram obtained in the traditional Stoner spin-fluctuation model for triplet superconductivity [7], where $T_s \to 0$ at $p_c$. However, when the strong-Ising anisotropy of the magnetization is taken into account a finite value of $T_s$ at $p_c$ can be attained in the model calculations [32].

The evolution of the upper critical field with pressure was investigated by resistivity measurements for $B \parallel a$ and $B \parallel c$ [17]. $B_{c2}^c$ is almost pressure independent, while $B_{c2}^a$ shows a remarkable enhancement upon approaching the critical pressure $p_c = 1.40$ GPa, with extrapolated values $B_{c2}^a(T\to 0)$ as large as 15 T. This demonstrates superconductivity is enhanced near the ferromagnetic quantum critical point. Measurements at $p = 1.66$ GPa show large values of $B_{c2}^a(0)$ persist in the paramagnetic phase, from which it has been inferred that $p$-wave superconductivity occurs at both sides of $p_c$.

## 8. Discussion

An important issue in the field of magnetic superconductors is whether superconductivity and magnetism are of bulk nature and coexist on the microscopic scale, or are confined to different parts of the sample because of phase separation on a macroscopic scale. Especially in the case of superconducting ferromagnets this is of major concern, as superconductivity and ferromagnetism form normally competing ground states. As regards UCoGe, solid evidence has been collected for the intrinsic coexistence of superconductivity and magnetism [1,16]. The thermodynamic signatures of the magnetic and superconducting phase transitions in the specific heat and thermal expansion of UCoGe show values characteristic for the bulk [1]. Moreover, the amplitude of the muon depolarization signal in the magnetic phase confirms bulk magnetism, which persists below $T_s = 0.6$ K [16]. The same conclusion was reached by $^{59}$Co-NQR measurements on poly and single-crystalline samples: below $T_C \approx 2.5$ K ferromagnetism and superconductivity are found to coexist on the microscopic scale [33]. From the temperature variation of the NQR spectrum, the authors conclude that the ferromagnetic phase transition is weakly first order. Notice, recent magnetization measurements on single crystals led to the claim that a magnetic field of the order of a few mT is needed to stabilize magnetic order [34]. This is at variance with the zero-field μSR [16] and NQR [33] results, and indicates metallurgy is an important issue (see section 2). The nuclear spin-lattice relaxation rate, $1/T_1$, in the ferromagnetic phase, extracted from the NQR experiments, decreases below $T_s$ due to the opening of the superconducting gap. Interestingly, two contributions to $1/T_1$ were found, *i.e.* terms proportional to $T$ and $T^3$. This has been interpreted to indicate the superconducting state is inhomogeneous. An appealing explanation for the inhomogeneous nature is the presence of a spontaneous vortex lattice [25], in which case the term linear in $T$ probes the non-superconducting regions of the sample, while the $T^3$ term probes the superconducting regions characterized by a line node in the superconducting gap function. This is in line with the interpretation of the μSR data (see section 5). UCoGe may be the first material in which a self-induced vortex state is realized. Small angle neutron scattering experiments and/or scanning squid probe techniques are needed to put this on firm footing.

Another important issue is the nature of the small ordered moment $m_0$. In analogy with other superconducting ferromagnets, it is natural to assume that the moment $m_0 = 0.07\ \mu_B$, deduced from the magnetization data, is due to U $5f$ electrons. Electronic structure calculations indeed predict a magnetic moment on the U site [35]. The calculated moment $\mu_U \sim 0.1\ \mu_B$ is small, due to an almost complete cancellation of the orbital $\mu_L^U$ and spin $\mu_S^U$ magnetic moment. However, the calculations predict the presence of a much larger moment $\mu_{Co} \sim 0.2$–$0.5\ \mu_B$ on the Co site as well. Recently, a polarized neutron diffraction study was conducted to solve the nature of the weak ferromagnetic moment [36]. Experiments carried out on a single-crystalline sample for $B \parallel c$ reveal that in low magnetic fields the ordered moment is predominantly located at the U moment. Thus ferromagnetic order is due to the $5f$ electrons. This is supported by the zero field μSR [16] and NQR [33] data. However, in a magnetic field the situation changes: the ordered moment grows to $\mu_U \sim 0.3\ \mu_B$ in a field of 12 T ($B \parallel c$) and, most remarkably, induces a substantial moment $\mu_{Co} \sim 0.2\ \mu_B$ on the Co atom, directed antiparallel to $\mu_U$. Such an anomalous polarizability of the Co $3d$ orbitals is unique among uranium intermetallics [21] and reflects the proximity to a magnetic instability of UCoGe in zero field.

The enhancement of superconductivity in UCoGe near the ferromagnetic quantum critical point provides an important clue that critical ferromagnetic spin fluctuations stimulate *p*-wave superconductivity. The condensation into spin-triplet Cooper pairs is in line with symmetry group considerations for superconducting ferromagnets with orthorhombic crystal symmetry [29]. Under the constraint of a large spin-orbit coupling and a sufficiently large exchange splitting, equal-spin pairing results in two-band superconductivity with gaps $\Delta_{\uparrow\uparrow}$ and $\Delta_{\downarrow\downarrow}$. Only two superconducting gap-structures are possible. By taking the ordered moment $m_0$ directed along the $z$ axis (as for an uniaxial ferromagnet), the gap has zeros (nodes) parallel to the magnetic axis ($k_x = k_y = 0$) or a line of zeros on the equator of the Fermi surface ($k_z = 0$). Accurate measurements of the electronic excitation spectrum in the superconducting state on high-purity single crystals are needed to discriminate between these two cases.

Since $p$-wave superconductivity is extremely sensitive to scattering at non-magnetic impurities and defects [7,37], a necessary condition for triplet pairing is a ratio of the mean free path over the coherence length $\ell/\xi > 1$. As mentioned above (section 6), our single crystals are sufficiently clean and we calculate $\ell/\xi \approx 7$. The sensitivity of superconductivity to the reduction of the mean free path has been investigated by doping UCoGe with Si [38]. Ac-susceptibility and resistivity measurements, carried out on a series of polycrystalline UCoGe$_{1-x}$Si$_x$ samples, show that superconductivity and ferromagnetism are progressively depressed with increasing Si content and simultaneously vanishes at a critical concentration $x_{cr} \cong 0.12$. Since the $RRR$ value rapidly drops with doping, and concurrently $\ell$ decreases, it is surprising triplet superconductivity survives till ~12 at.% Si. This would require a strong doping-induced reduction of $\xi$ as well. In the case of UCoGe$_{1-x}$Si$_x$, however, the defect-driven depression of $T_s$ is partly compensated by its increasing due to chemical pressure. Also, upon approach of the ferromagnetic quantum critical point, ferromagnetic spin fluctuations will promote triplet superconductivity even stronger.

The superconducting phases of UCoGe under pressure, labelled $S_1$ and $S_2$ in the $p$-$T$ diagram (Fig. 8), can be discriminated in close analogy to the familiar superfluid phases of $^3$He [39]. The state $S_1$ in the ferromagnetic phase breaks time reversal symmetry and is equivalent to the non-unitary $A_2$ phase of $^3$He (*i.e.* the $A$ phase of $^3$He in magnetic field), which is a linear combination of the equal-spin pairing states $|S_z = 1, m = 1\rangle$ and $|S_z = -1, m = 1\rangle$ with different population. The large upper-critical field values observed for state $S_2$ provide solid evidence it is a spin-triplet state as well. Model calculations [39] predict it is a unitary triplet state, which does not break time reversal symmetry. In this sense, $S_2$ is equivalent to the planar state of $^3$He, which is an equally weighted superposition of the two states $|S_z = 1, m = -1\rangle$ and $|S_z = -1, m = 1\rangle$. UCoGe is unique as regards its response to pressure, as it is the only superconducting ferromagnet for which superconductivity persists above the critical pressure for suppression of ferromagnetism. Very similar $p$-$T$ phase diagrams have been obtained by other research groups on poly- and single-crystalline UCoGe samples [40], which demonstrates the pressure response reported in Fig. 8 is a robust property.

Recently, a second route to quantum criticality in UCoGe has been explored, namely magnetic field tuning. When carrying out a high-field magnetotransport study on high quality single crystals ($RRR \approx 30$) at ambient pressure, Aoki and co-workers [30] made a remarkable observation: $B_{c2}^b(T)$ is strongly enhanced and shows an unusual S-shaped curve which extrapolates to the large value of 20 T when $T \to 0$. For $B \parallel a$ even larger upper critical field values are attained: $B_{c2}^a(T)$ shows an unusual upward curvature and extrapolates to 30 T for $T \to 0$. A key ingredient in the magnetotransport experiment is precise tuning of the magnetic field along the orthorhombic axes of the crystal. A misalignment of a few degrees inhibits the observation of these phenomena, which most likely explains the much lower $B_{c2}^{a,b}$ values reported in Ref.15 (see Fig. 7). The field-reinforced superconducting phase for $B \parallel b$ appears to be connected to the depression of the Curie temperature in large magnetic fields (with a critical field $B_c \sim 16$ T for $T \to 0$) [30,41]. Compelling evidence for a close link between critical spin fluctuations at $B_c$ and superconductivity is obtained by analyzing the resistivity data in the normal state. The underlying idea is that critical spin fluctuations, which are the source of the pairing interaction, give rise to an enhanced quasiparticle mass, $m^*$, which can be probed via the Fermi liquid term in the resistivity, $\rho \sim AT^2$. This can be made quantitative by use of a simple McMillan-like relation between $T_s$ and $m^* \propto \sqrt{A}$ for ferromagnetic superconductors, recently proposed in Ref. 11. As expected, for $B \parallel b$ the transport coefficient $A$ strongly increases with field and shows a pronounced maximum in the field range of reinforced superconductivity [30]. A similar relation was recently established for URhGe [10,11], where field re-entrant superconductivity is due to critical spin fluctuations associated with a spin-reorientation process which is induced in a large magnetic field ($B \parallel b$) of 12 T [42].

## 9. Conclusions

In this review we have presented the thermal, magnetic, and transport properties of the superconducting ferromagnet UCoGe. The data obtained on high-quality poly- and single-crystalline samples show the unusual coexistence of ferromagnetism and superconductivity is robust on the macroscopic and microscopic scale. Notably, the absence of Pauli limiting in the $B-T$ phase diagrams and the enhancement of superconductivity near the magnetic quantum critical point in the $p-T$ diagram provide evidence for triplet superconductivity mediated by critical ferromagnetic spin fluctuations. UCoGe offers a unique possibility to further unravel the intimate link between ferromagnetism and triplet superconductivity. Especially, the two routes to quantum criticality – pressure tuning and magnetic field tuning – make it an unrivalled laboratory tool to probe spin-fluctuation mediated superconductivity. We expect in the near future measurements of the electronic and magnetic excitation spectra in the superconducting and magnetic phases will reveal crucial information on the superconducting gap structure and pairing mechanism. These experiments inevitably should be performed on high-purity single crystals, which calls for a strong commitment to further improve the sample preparation process. UCoGe is a unique test-case material for addressing the central issue of how a ferromagnetic superconductor accommodates an intrinsic internal magnetic field. It may be the first material to reveal proof for the existence of the long-searched-for spontaneous vortex phase.

**Acknowledgement**

The authors are grateful to T. Gortenmulder, D.E. de Nijs, A. Hamann, T. Görlach, H. v. Löhneysen, C. Baines, D. Andreica and A. Amato for their help at various stages of the research. We thank D. Aoki, V.P. Mineev and A.D. Huxley for helpful discussions. This work is part of the research programme of the Foundation for Fundamental Research on Matter (FOM), which is financially supported by the Netherlands Organisation for Scientific Research (NWO), and of the EC $6^{th}$ Framework Programme COST Action P16 ECOM.

FIGURES

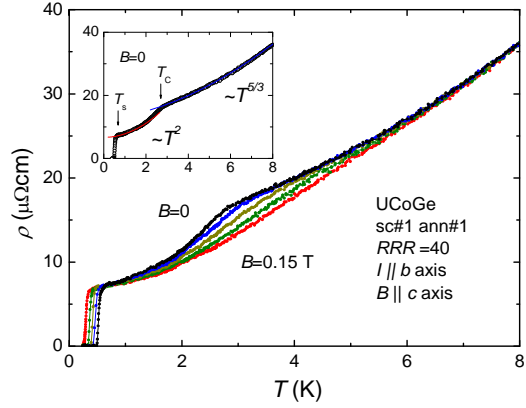

Fig.1 Temperature variation of the resistivity of annealed single-crystalline UCoGe for a current $I \parallel b$ in zero magnetic field (upper curve) and in applied fields $B \parallel c$ of 0.02, 0.05, 0.10 and 0.15 T (lowest curve). The residual resistivity $\rho_0 = 7$ μΩcm and $RRR = 40$. Inset: Resistivity in zero field, where the solid lines represent fits to $\rho \sim T^2$ and $\sim T^{5/3}$ in the temperature ranges below and above $T_C$, respectively. Arrows indicate the Curie temperature $T_C = 2.8$ K and the onset temperature for the superconducting transition $T_s^{onset} = 0.6$ K.

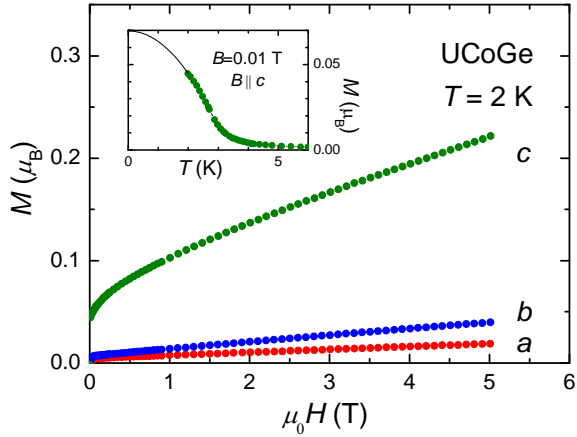

Fig.2 Magnetization of UCoGe for fields along the $a$, $b$ and $c$ axis at $T= 2$ K [15]. Ferromagnetic order is uniaxial with $m_0$ pointing along the $c$ axis. The inset shows $M(T)$ for $B = 0.01$ T along $c$. In the limit $T \to 0$ $m_0 = 0.07$ μ$_B$.

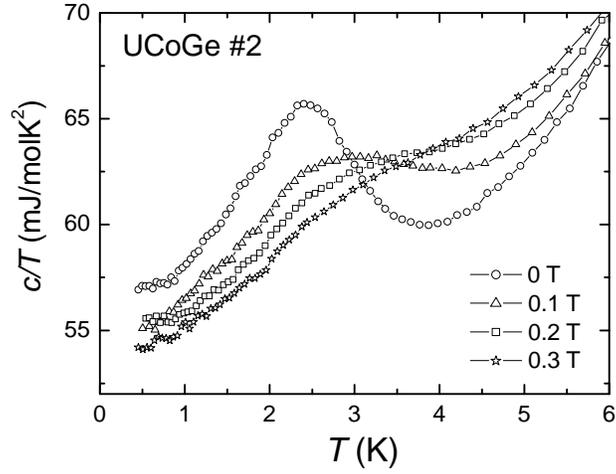

Fig. 3 The temperature dependence of the specific heat of polycrystalline UCoGe (sample #2) plotted as $c/T$ versus $T$ in fields of 0, 0.1, 0.2 and 0.3 T [23].

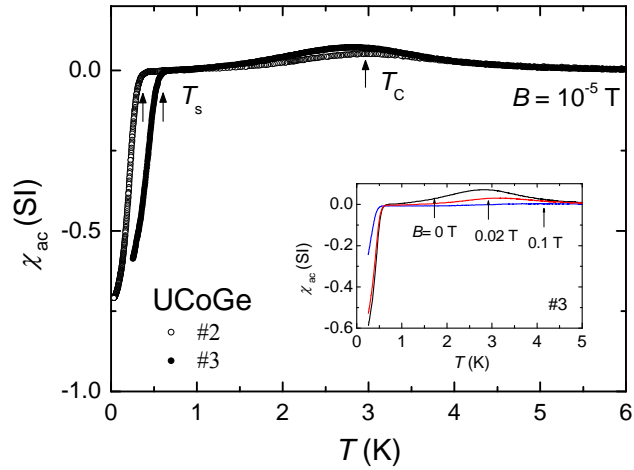

Fig. 4 Temperature dependence of the ac-susceptibility $\chi_{ac}$ in polycrystalline UCoGe (samples #2 and #3) [1,23]. Arrows indicate $T_C$ and $T_s$. Inset: The ac-susceptibility of sample #3 measured in fields of 0, 0.02 and 0.1 T.

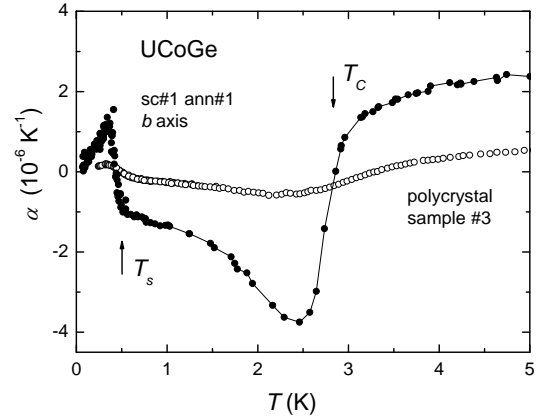

Fig. 5 Temperature variation of the coefficient of linear thermal expansion of UCoGe. Open circles: polycrystal (sample #3) [1]. Closed circles: single-crystal (sc #1, ann #1) *b* axis.

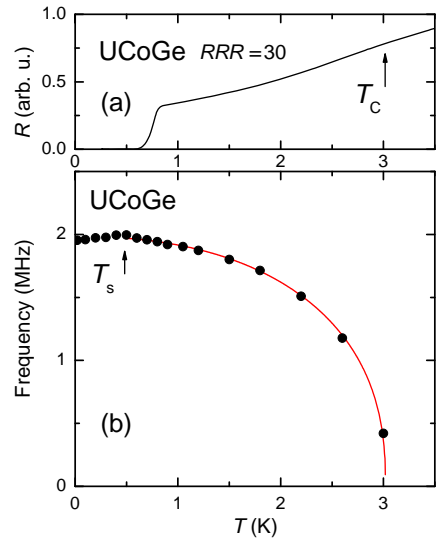

Fig. 6 (a) Resistivity versus temperature of polycrystalline UCoGe. (b) Spontaneous muon precession frequency $\nu(T)$ of UCoGe. The solid line represents a fit to a phenomenological order parameter function (see text). Notice the 2% decrease of $\nu(T)$ below the bulk $T_s$ of 0.5 K. Figure adapted from Ref.16.

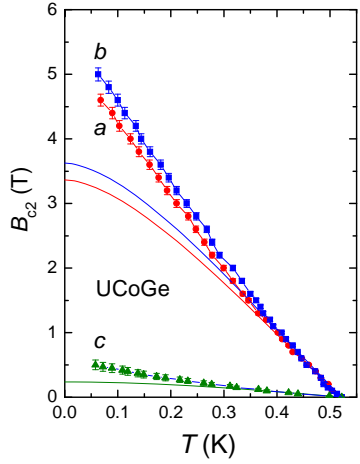

Fig. 7 Temperature dependence of the upper critical field of single-crystalline UCoGe for $B$ along the principal axes [15]. The solid lines show the calculated dependence for a superconducting gap function with axial (along $c$ axis) and polar symmetries (along $a$ and $b$ axis).

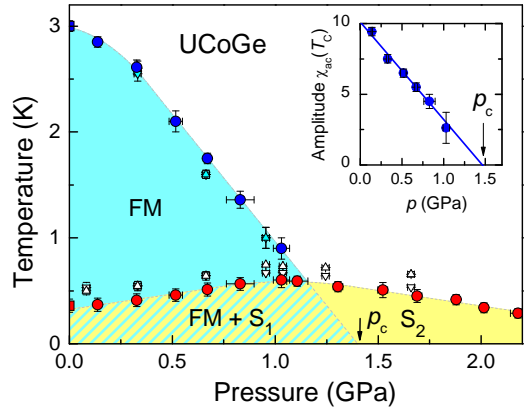

Fig. 8 Pressure temperature phase diagram of UCoGe [17]. Ferromagnetism (FM) - blue area; superconductivity (SC: $S_1$, $S_2$) - yellow area. $T_C(p)$ extrapolates to a FM quantum critical point at $p_c = 1.40\pm0.05$ GPa. SC coexists with FM below $p_c$ - blue-yellow hatched area. Symbols: closed blue and red circles $T_C$ and $T_s$ from $\chi_{ac}(T)$; blue and white triangles $T_C$ and $T_s$ from $\rho(T)$ (up triangles $I \parallel a$, down triangles $I \parallel c$); closed blue and red squares $T_C$ and $T_s$ at $p=0$ taken from a polycrystal. Inset: Amplitude of $\chi_{ac}(T)$ at $T_C$ as a function of pressure. The data follow a linear $p$-dependence and extrapolate to $p_c = 1.46\pm0.10$ GPa.